\documentclass[useAMS]{mn2e}
\usepackage{psfig,epsfig,natbib_sro}

\title[High Resolution Calculations of Merging Neutron Stars I]{High Resolution Calculations of Merging Neutron Stars I: 
Model Description and Hydrodynamic Evolution}
\author[S. Rosswog and M.B. Davies]
       {Stephan Rosswog, Melvyn B. Davies \\
           Department of Physics and Astronomy,University of Leicester, 
           LE1 7RH, Leicester, UK\\}
\date{Accepted 2002 January 31.
      Received 2002 January 17;
      in original form 2001 September 4}

\pagerange{\pageref{firstpage}--\pageref{lastpage}}
\pubyear{2001}

\begin{document}
\newcommand{\newblock}{}
\maketitle

\begin{abstract}

We present the results of 3D high resolution calculations  of the last 
inspiral stages and the final coalescence of neutron star binary systems.
The equations of hydrodynamics are solved using the smoothed particle
hydrodynamics method with up to $10^6$ particles.
Using Newtonian gravity, but adding the forces emerging from the emission of 
gravitational waves, we focus on the impact of microphysics on the dynamical 
evolution of the merger. Namely we use a new equation of state based on the
relativistic mean field approach of \cite{shen98a,shen98b}.
Neutrino emission of all flavours, the resulting cooling
and the change in the electron fraction are accounted for with a detailed 
leakage scheme.\\
The new equation of state is substantially stiffer than the Lattimer-Swesty
equation of state that has been used in previous investigations. This leads
the system to become dynamically unstable at a separation as large as 3.3
stellar radii, where the secular orbital decay undergoes a transition towards
a dynamical ``plunge'' of the binary components towards each other. 
As soon as the stars come into contact a 
Kelvin-Helmholtz-instability forms at the interface of both stars.  
The peak temperatures are found in the vortex rolls that form during this 
process. We generally find slightly lower temperatures than
previously found using the Lattimer-Swesty equation of state.\\
The central object is surrounded by a very thick disk that shows  cool 
equatorial flows: inflows in the inner parts of the disk and outflow further
 out. The cool inflows become shock heated in the innermost parts of the disk
and lead to an outflow of hot material in the vertical direction.
The temperatures in the disk have typical values of 3-4 MeV, lower than the 
temperatures found in previous investigations using the Lattimer-Swesty 
equation of state. These conditions allow for the existence
of heavy nuclei even in the inner parts of the disk, we find typical mass
fractions of $\sim 0.1$, which is enough for scattering off heavy
nuclei to be the dominant source of neutrino opacity.\\
The central object formed during the coalescence shows a rapid, differential 
rotation with periods of $\sim 2$ ms. Although a final conclusion on this 
point is not possible from our basically Newtonian approach, we argue that 
the central object will remain stable without collapsing to a black hole,
at least on the simulation time scale of 20 ms, mainly stabilized by  
differential rotation. The massive, differentially rotating central object
is expected to  wind up initial magnetic fields to enormous field strengths of
$\sim 10^{17}$ G (Duncan $\&$ Thompson 1992) and may therefore have important 
implications for this event as a central engine of Gamma Ray Bursts.
\end{abstract}


\section{Introduction}
The final stage of the inspiral and the subsequent coalescence of a neutron 
star binary system is among the prime candidates for ground based 
gravitational wave detection
via the laser interferometers currently under construction such as LIGO 
\citep{abramovici92}, TAMA \citep{kuroda97}, VIRGO \citep{bradaschia90} and GEO 
\citep{danzmann95}. The neutron star merger scenario could also provide the
energy reservoirs to power cosmological gamma ray bursts (GRB;
\cite{paczynski86,eichler89}) and is in addition one of the two most discussed
production sites of the heavy, rapid neutron capture elements 
\citep{lattimer74,lattimer76,symbalisty82,eichler89,rosswog99,freiburghaus99b}.
Whether this scenario can fulfil all these promises or not, neutron star 
binary systems are
known to exist and to inevitably coalesce \citep{taylor94}. This fact alone
 - apart from all further promises - justifies a careful study of the
coalescence process and its possible implications.\\
The coalescence is an intrinsically three-dimensional 
phenomenon and therefore analytical guidance is rare although very welcome and
one has to resort to large scale computations. Additional complications
arise from the fact that there is almost no field of astrophysics that does 
not enter at some stage during the coalescence process: the last stages and 
the 
merger are certainly dominated by strong-field general relativistic gravity, 
the neutron star material follows the laws of hydrodynamics, particle physics
enters via possible condensates of ``exotic'' matter in the high-density 
interiors of the neutron stars and the copiously produced neutrinos in the 
hot and dense neutron star debris, questions concerning
element formation require detailed information on nuclear structure and 
reactions (often far from stability) to be included and also magnetic 
fields might play a decisive role since they may, via transport of angular
momentum, determine whether and/or when the central, coalesced object collapses
into a black hole.\\
Due to this complexity current investigations follow one of two
``orthogonal'' lines: either ignoring microphysics, resorting to the simplest
 equations of state (EOS), polytropes, and thereby 
focusing on
solving the complicated set of general relativistic fluid dynamics equations
(or some approximation to it) or, along the other line, using 
Newtonian self-gravity of the fluid and 
investigating the influences of detailed microphysics and relating the 
merger event to astrophysical phenomena.\\
Many groups have performed 3D calculations of the merger scenario. 
The first work, 
using a polytropic EOS and Newtonian gravity, was performed by 
\citet[and references therein]{oohara97}  using Eulerian finite difference 
codes. \citet{rasio92,rasio94,rasio95},  
\citet{zhuge94,zhuge96} and  \citet{davies94} have used the 
smoothed particle hydrodynamics method (SPH) to further explore various aspects
of the scenario.\\
The strong-field gravity aspect of the event was explored again by
 \citet[and references therein]{oohara97} using Post-Newtonian 
hydrodynamics in Eulerian formulation. Recently, Post-Newtonian calculations
 have also been performed using SPH by  \citet{ayal01}, 
\citet{faber00} and \citet{faber01}. Several groups have worked on 
general relativistic formulations where the metric has been treated in the 
conformal flatness approximation 
\citep{wilson95,oohara97,baumgarte97,oechslin01}. Recently, much progress has 
been made towards a stable implementation of the fully relativistic equations
of hydrodynamics \citep{shibata99,shibata00}.\\
The line relying on (basically) Newtonian gravity and investigating the 
microphysics of the event has been explored by 
\citet{ruffert96,ruffert97a} and \citet{ruffert01} using a grid  
code based on the piecewise parabolic method (PPM) and by 
\citet{rosswog99,rosswog00} using SPH.\\
In this paper we want to continue to explore  the impact 
of the microphysics on the outcome of the coalescence. We use a new equation
of state that overcomes previous restrictions in the temperature and density.
Since the high density regime of the nuclear physics is only poorly known
to date this work will unfortunately not provide the final answer, but 
further broaden our understanding of the large effects that can be expected 
from the nuclear physics input.
The calculations to be described below were performed on parallel computers
allowing for an unprecedented hydrodynamic resolution.\\
The paper is organized as follows.
In Section 2 we describe the basic ingredients like the hydrodynamics or the 
equation of state of our model. Section 3 describes the initial conditions
of our calculations, the overall hydrodynamic evolution, and then focuses
on aspects concerning the central object and the debris, referred to as 
``disk''. The summary and a discussion of the results are provided in 
Section 4.

\section{Model}

%%%%%%%%%%%%%%%%%%%%%%%%%%%%%%%%%
%       hydrodynamics           %
%%%%%%%%%%%%%%%%%%%%%%%%%%%%%%%%%
\subsection{Hydrodynamics}
To follow the dynamical evolution of the neutron star fluid we use a 
Lagrangian particle scheme, the so-called smoothed particle hydrodynamics
method (SPH; \cite{benz90a,monaghan92}).
Since it is independent of any prescribed geometry (e.g. grid) it is perfectly
suited to handle the intrinsically three-dimensional merger process. In 
addition, the Lagrangian nature of the scheme makes it easy to carefully
track the evolution of interesting portions of the fluid 
(e.g. possible ejecta). Voids are treated in a natural way (i.e. no particles)
and do not present any difficulty for the method,
the interesting parts of the simulation do not have to be embedded in
an artificial background medium like in other methods (e.g. PPM). The use of a
background medium leads to further difficulties like emerging (artificial)
shock waves at the stellar surfaces that have to be treated by additional 
remedies \citep{ruffert01}. If the corresponding parameters are not chosen
carefully, the artificial medium may also lead to a damping of (physical) 
oscillations \citep{font00}. 

Since SPH is well-known we will only describe the basic ingredients of our
code. The basic set of equations may be found in \cite{benz90a}.
Rather than the ``standard'' form of {\em artificial viscosity}
 (AV; \cite{monaghan83})
we use a hybrid method that profits from two improvements: the viscosity 
parameters are time dependent \citep{morris97} and have non-negligible
values only in the presence of shocks. Additionally, a switch is applied that 
suppresses spurious forces in pure shear flows \citep{balsara95}. This hybrid
method \citep{rosswog00} has been shown to resolve shocks with the same 
accuracy as the standard formulation of AV, but exhibits much better 
behaviour in shear flows.
In test calculations of differentially rotating stars the viscous time scales
$\tau_{visc,i}= v_i / \dot{v}_{i,visc}$ obtained with the new scheme were two 
orders of magnitude longer than those from the standard form of AV. For 
details and test calculations we refer to \cite{rosswog00}.\\
Since it is a widespread belief that SPH is always extremely viscous by nature,
we want to measure the effective ``alpha-viscosity'', $\alpha_{ss}$, 
defined by \cite{shakura73}
\begin{equation}
\nu = \alpha_{ss} c_s H,\label{shakura}
\end{equation}
where $\nu$ is the kinematic viscosity, $c_s$ the sound velocity and $H$ the
disk thickness. In 3D the SPH bulk viscosity parameter, $\alpha$ 
(see e.g. \cite{monaghan92}), can be related to the kinematic viscosity via 
(\cite{murray95}) 
\begin{equation}
\nu = \frac{\alpha}{10} c_s h,\label{james}
\end{equation}
where h is the smoothing length. Due to the modifications in our hybrid 
AV-scheme we use $\tilde{\alpha}_i= \alpha_i \cdot f_i$ in the above formula
rather than the bulk viscosity parameter of particle $i$, $\alpha_i$. 
Here $f_i$ is the Balsara-factor implemented in our 
code,
\begin{equation}
f_i= \frac{|\nabla\cdot\vec{v}|_i}{|\nabla\cdot\vec{v}|_i + 
|\nabla \times \vec{v}|_i + \eta c_{s,i}/h_i}
\end{equation}
and $\eta=10^{-4}$. This factor suppresses AV in pure shear flows 
($f_i \rightarrow 0$ for $|\nabla \times \vec{v}| \gg |\nabla\cdot\vec{v}|$) where it is unwanted, and $f_i \rightarrow 1$ in the case of pure shocks ($|\nabla \times \vec{v}| \ll |\nabla\cdot\vec{v}|$).
Equating (\ref{shakura}) and (\ref{james}) one finds
\begin{equation}
\alpha_{ss,i} = \frac{\tilde{\alpha}_i}{10} \frac{h_i}{H},\label{a_SPH}
\end{equation}
i.e. apart from the numerical value $\tilde{\alpha}_i$ it is the resolution of 
the disk that determines the effective alpha-viscosity $\alpha_{ss}$. It 
should, however, be noted that this relation is only approximate since the 
effects of
AV and the $\alpha$-viscosity prescription are not necessarily exactly 
the same and due to the way the Balsara-factor is taken into account.
To obtain numerical values we insert the average thickness of the central 
object 
and disk of run B (see below; Table \ref{runs}) into equation (\ref{a_SPH}).
By averaging according to
\begin{equation}
\langle \alpha \rangle_{ss}= \frac{\sum_i m_i \alpha_{ss,i}}{\sum_i m_i},
\end{equation}
where $m_i$ are the particle masses, we find
\[
\langle \alpha \rangle_{ss}= \left\{ \begin{array}{r@{ }l}
                         &6 \cdot 10^{-4}  \quad \mbox{all particles}\nonumber\\
                         &4 \cdot 10^{-3}  \quad \mbox{debris}\nonumber\\
	                 &3 \cdot 10^{-5}  \quad \mbox{central object} \nonumber
                                     \end{array}
                             \right. \]
For details of concerning ``debris'' and ``central object'' see below.
These values are very low in comparison with the values that are derived from 
observations of cataclysmic variables, $\alpha\sim 0.1$, and it is certainly 
possible that the physical viscosity of a hot neutron star debris disk is
much higher than the values of the above determined numerical viscosity.\\
The (Newtonian) forces of the {\em self-gravity} of the fluid are efficiently 
calculated using a binary tree \citep{benz90b}.
Since the computationally most expensive part of the code is the evaluation of 
gravitational forces we implemented an {\em integrator} for the set of 
differential equations that only needs one 
force evaluation per time step. We decided to implement an Adams-Bashforth 
method which is third order accurate in time. This allows for a very accurate
conservation of energy and angular momentum even with relative large step 
sizes. In a calculation with $\sim$ 100 000 particles these quantities are 
conserved to a few parts in $10^{4}$ over $\sim$ 10000 time steps. For reasons 
of comparison: other schemes \citep{ruffert01} lose up to 10 $\%$ of the
 total angular 
momentum by numerical artifacts despite excellent energy conservation.\\
The amount of noise present in SPH depends, of course, on the particle 
distribution. For the worst case, a stochastic particle distribution,
it scales like $1/\sqrt{N_N}$, where $N_N$ is the number of neighbours 
a particle interacts with. Therefore while increasing the total particle
number, $N$, one should also increase the number of neighbours in order 
to reduce the noise level. On the other hand one is not interested in 
too large an 
increase of the neighbour number and therefore the smoothing length, $h$,
since it would  compromise the spatial resolution and the numerical error; 
remember that the SPH-equations are accurate to order $O(h^2)$. 
Therefore, $N_N$ should
increase slower than N. In the calculations presented here
typical neighbour numbers are 100-120.\\
Our whole code is parallelized for use on shared memory machines using 
OpenMP. For particle numbers above $\sim$
400 000 we find an almost linear speedup for up to 120 processors.

%%%%%%%%%%%%%%%%%%%%%%%%%%%%%%%%
%      GW-backreaction         %
%%%%%%%%%%%%%%%%%%%%%%%%%%%%%%%%
\subsection{Gravitational Wave Back-reaction}
The forces emerging from the emission of gravitational waves that drive the
binary towards coalescence are treated in the
point mass limit of the quadrupole approximation and are applied until 
the stars come into contact. Since each particle in a star is subject to
the same back-reaction force the conservation of the fluid 
circulation is guaranteed. For a further discussion of this approach 
we refer to \citet{rosswog99}.

%%%%%%%%%%%%%%%%%%%%%%%%%%%%%%%%
%	     EOS               %
%%%%%%%%%%%%%%%%%%%%%%%%%%%%%%%%
\subsection{Equation of State}
\begin{figure}
%    \hspace*{0cm}\psfig{file=Pressure_ext.eps,width=8cm,angle=-90}
    \hspace*{0cm}\psfig{file=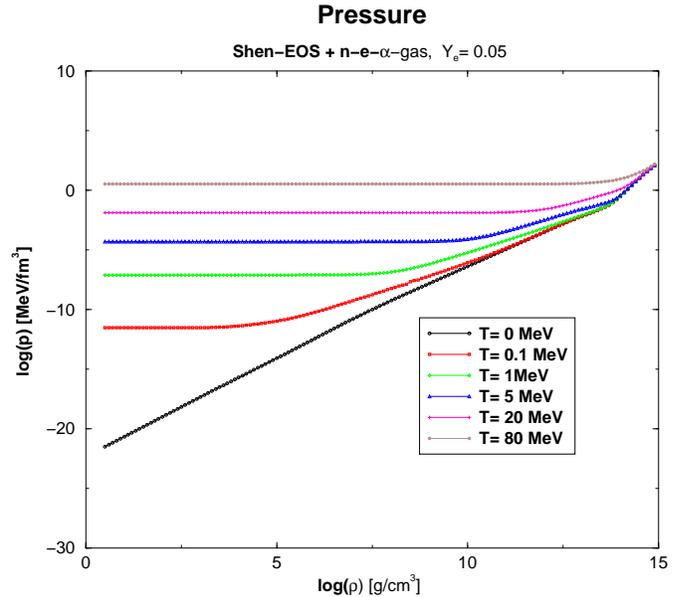,width=8cm,angle=-90}
    \caption{\label{press_ext} Pressure as a function of temperature for our 
EOS (baryonic part from Shen et al. + photons + electron-positron pairs,
extension with a gas consisting of neutrons, alpha particles,  
electron-positron pairs and photons). The transition between the original
EOS and the extension lies at $\rho\approx 10^5$ g/cm$^3$.}
\end{figure}

The equation of state (EOS) is one of the most crucial ingredients of a 
neutron star simulation and the microscopic behaviour of matter decisively 
determines the overall, macroscopic evolution of the system (compare, for 
example, the calculations using a polytrope versus those with the nuclear
EOS of \citet{lattimer91} [LS-EOS] in \cite{rosswog99}).
Since the stiffness of the EOS varies over a wide range as a function of
density, temperature and composition (see below) and the release of nuclear
binding energy when nucleons form nuclei has important dynamical consequences,
a simple polytropic EOS is only a poor approximation of the involved 
microphysics.
All neutron star merger calculations that used a nuclear equation of state
\citep{ruffert96,ruffert97a,ruffert01,rosswog99,rosswog00}
relied on the Lattimer-Swesty-EOS. This EOS,
however, has been designed for the use in supernova-calculations and therefore
suffers from some deficiencies in our context: the electron
fraction, $Y_e$, is restricted to values above $\sim 0.04$, temperatures 
to values above $\sim 10^9$ K and densities above $10^7$ gcm~$^{-3}$. 
The electron fraction in a neutron star in 
$\beta$-equilibrium, however, dips down to values of $\sim 0.01$ near the neutron
star surface (see below) and temperatures in old neutron stars are expected 
to be negligible with respect to typical nuclear energies; viscous heating 
during the inspiral is expected to increase the temperatures to only 
$\sim 10^8$ K \citep{lai94c}. Therefore, once outside the EOS-range, these 
investigations either used
extrapolation \citep{ruffert96} or kept the EOS-quantities at 
the boundary values \citep{rosswog99}.\\
In this work we use an EOS  based on the tables of  
\citet{shen98a,shen98b} that overcomes all above mentioned restrictions, 
in the electron fraction as well as in temperature and density.
Shen et al. follow a  relativistic mean field (RMF) approach
which reproduces the basic features of more complicated 
relativistic Dirac-Br\"uckner-Hartree-Fock calculations.
The starting point is a relativistically covariant Lagrangian that
contains, apart from the nucleon fields, the scalar $\sigma$, and the 
$\omega$ and $\rho$ vector mesons.  
For the  $\sigma$ and $\omega$ meson fields
non-linear terms are included, which are essential to reproduce quantitative
properties of nuclei (e.g. the compression modulus) correctly. 
The parameter set (TM1, \cite{sugahara94}) is chosen in a way that 
experimental data of finite nuclei in the ground as well as in excited states
are satisfactorily reproduced.
With this parameter set the compression modulus, $K_0$, has a value
of 281 MeV, whereas in the Lattimer-Swesty EOS in \cite{ruffert96},
\cite{rosswog99} and all comparisons in this paper $K_0= 180$ MeV has 
been used.
Since many-body interactions at high densities are only poorly constrained to
date, no effort is made to include more ``exotic'' physics such as hyperons, 
various mesons or quark matter in the high density regime.\\
At densities above $\sim 1/3 \; \rho_{nuc}$ protons and neutrons 
form a homogeneous ``nucleon fluid'', 
below this density matter
may become inhomogeneous, i.e. the presence of nuclei may become 
energetically 
favorable. This phase is modelled using the Thomas-Fermi approximation.
Matter is assumed to consist of a mixture of nucleons, alpha-particles and 
spherical nuclei arranged on a lattice. The heavy nucleus (representative
of a distribution of heavy nuclei) is assumed to be centered in a charge 
neutral cell consisting of vapor of the neutrons, protons and alpha particles.
Above $\rho \approx 10^{10}$ g cm~$^{-3}$ the nucleons are treated by 
the RMF theory, below this density they are assumed 
to form a Maxwell-Boltzmann gas. The alpha-particles are treated as a 
non-interacting Boltzmann gas, their occupied volume is accounted for
in the calculation of the free energies. The density distribution in
the Wigner-Seitz cell is parametrized and the free parameters are determined by
minimizing the free energy density with respect to the densities of all 
ingredients.\\
Shen et al. only provide the baryonic part of the EOS. Their tables 
range from 0 to 100 MeV in temperature, 0 to $\sim$ 0.56 in $Y_e$ and
5.1 to 15.4 in $\log(\rho$) ($\rho$ in cgs-units).
We therefore add the contributions from photons and electron-positron pairs to 
the baryonic components. For the electron-positron pairs we use the code
of \cite{timmes99}.
Apart from disregarding interactions, which is
perfectly justified at the high densities of interest, the 
electron-positron-pairs are treated in full generality without any 
approximation.
In the low density regime, 
$\rho < 10^{5}$ g cm~$^{-3}$, we extend the EOS with a gas consisting of 
neutrons, alpha-particles, electron-positron pairs and photons. 
The smooth transition is demonstrated in Figure \ref{press_ext}.
\begin{figure}
%    \hspace*{0cm}\psfig{file=LS_Shen_comparison_close.eps,width=8cm,angle=-90}
    \hspace*{0cm}\psfig{file=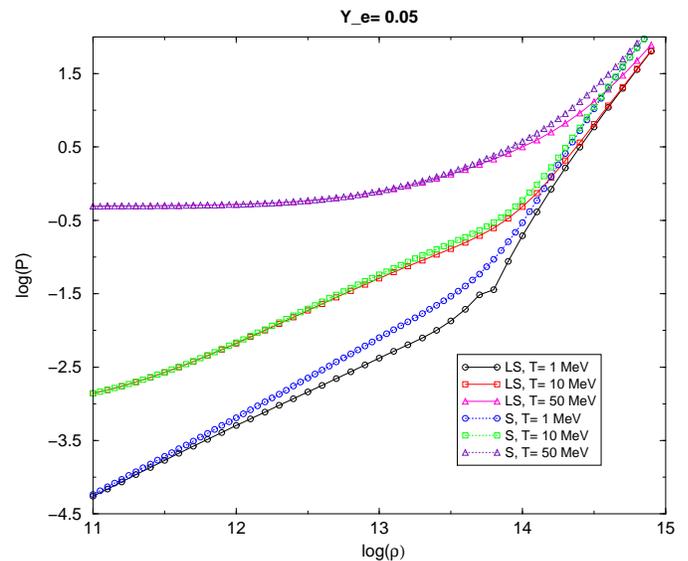,width=8cm,angle=-90}
    \caption{\label{LS_Shen_comp} Comparison of the pressures provided by
the Lattimer-Swesty and the Shen et al. EOS as a function of density 
(cgs-units).}
\end{figure}
The EOS  described above is tabulated with 32 entries in temperature, 72 in
electron fraction and 151 in density. We store the following quantities:
total pressure, internal energy, difference in the nucleonic chemical 
potentials $\hat{\mu}$, sound velocity, entropy, mass fractions of 
protons ($x_p$), alpha particles ($x_{\alpha}$) and the heavy nucleus ($x_h$),
 the proton to nucleon ratio, $Z/A$, 
and the nucleon number $A$ of the heavy nucleus. The neutron mass fraction is 
obtained as $x_n= 1 - (x_p + x_{\alpha} + x_h)$.\\
In Fig. \ref{LS_Shen_comp} we show a comparison between the pressures provided
by both the LS-EOS and the EOS of Shen et al. Below $10^{11}$ gcm~$^{-3}$ the 
results are practically identical, at higher densities the new EOS is 
substantially stiffer than the LS-EOS. For rather low temperatures ($T=1$ MeV)
we find large differences in the density range from $10^{12}$ to $10^{14}$
gcm~$^{-3}$, which is the density regime that is crucial for the torus that 
forms
around the central object, see below. For even higher densities the pressure
curves for different temperatures converge indicating the diminishing relevance
of thermal pressure contributions. Again, the Shen et al. EOS yields 
considerably higher pressure values.\\
The Shen-EOS also yields larger values for 
$\hat{\mu}$. Since the neutrino-less $\beta$-equilibrium value of the electron 
chemical potential is given by $\bar{\mu}_e= \hat{\mu} -Q$, $Q$ being the 
neutron proton mass difference, and $Y_e$ is a monotonic function of 
$\bar{\mu}_e$ the Shen-EOS yields higher values for $Y_e$ than the LS-EOS.

%%%%%%%%%%%%%%%%%%%%%%%%%%%%%%%%
%       neutrino physics       %
%%%%%%%%%%%%%%%%%%%%%%%%%%%%%%%%
\subsection{Neutrino Physics}
The emission of neutrinos provides a very efficient cooling mechanism for hot
neutron star matter and the related electron and positron captures change the 
electron fraction $Y_e$ which in turn determines the matter composition.
We include the most important neutrino reactions where we ensure, via effective
rates, that only the amount of neutrinos is produced that is actually
able to leave the 
dense surrounding material. Our scheme takes careful account of the energy 
dependence of the neutrino opacities by integrating over the neutrino 
distributions.\\
For the emission processes we include  
electron captures (EC)
\begin{equation} 
     e^- + p \rightarrow n + \nu_e  
\end{equation}
and positron captures (PC)
\begin{equation}
     e^+ + n \rightarrow p + \bar{\nu}_e 
\end{equation}
which produce electron flavour neutrinos
and the pair producing reactions,
pair annihilation
\begin{equation} 
     e^- + e^+ \rightarrow \nu_i + \bar{\nu}_i 
\end{equation}
and plasmon decay
\begin{equation} 
     \gamma \rightarrow \nu_i + \bar{\nu}_i. 
\end{equation}
Here $\bar{\nu}_i / {\nu}_i$ refer to anti-/neutrinos of all types.\\
For the opacities we include the dominant processes,
neutrino nucleon scattering 
   \begin{equation} \nu_i + \{n, p\} \rightarrow \nu_i + \{n, p\},
   \end{equation}
coherent neutrino nucleus scattering
   	\begin{equation}
        \nu_i + A \rightarrow \nu_i + A, 
   	\end{equation}
and neutrino absorption by nucleons
	\begin{eqnarray}
	\nu_e + n \rightarrow p + e^-\\
	\bar{\nu}_e + p \rightarrow n + e^+.
	\end{eqnarray}  
Note that this is the first time that the effects of scattering
off heavy nuclei have been accounted for in a neutron star merger calculation.
These are important whenever nuclei are present since the corresponding 
cross sections scale proportional to $A^2$ and $A\approx 80$ (see below).
For details of this transport scheme we refer to 
\cite{rosswog01c}, where the neutrino emission results are discussed in detail.

\section{Results}
%%%%%%%%%%%%%%%%%%%%%%%%%%%%%%%%
%    initial conditions        %
%%%%%%%%%%%%%%%%%%%%%%%%%%%%%%%%
\subsection{Initial Conditions}
\subsubsection{Neutron Star Masses}
We focus in  this investigation on equal mass
systems with 1.4 M$_{\odot}$ per star since all well-determined neutron star 
masses from radio binary pulsars are distributed
according to a  narrow distribution around 1.4 M$_{\odot}$ 
\citep{thorsett99}. 
For neutron stars in X-ray binary systems substantially higher masses have
been claimed: e.g. $1.8 \pm 0.2$ M$_{\odot}$ for CygX-2 \citep{orosz99},
$\sim 1.9$  M$_{\odot}$ for VelaX-1 \citep{vankerkwijk95} and $1.8 \pm 0.4$
 M$_{\odot}$ for 4U 1700-37 \citep{heap92}. In order to get an upper limit 
we additionally explore the case of twice 2.0  M$_{\odot}$ 
(see Table \ref{runs}).\\
It should be noted that the system dynamics 
is extremely sensitive to even small deviations from a mass ratio of unity.
This sensitivity to the mass difference has been explored previously, both 
using a polytropic \citep{rasio94} and a realistic EOS 
\citep{rosswog00,ruffert01} and will not be the subject of this investigation.

\subsubsection{Neutron Star Models}
\begin{figure}
%    \hspace*{0cm}\psfig{file=NSprofilesSHEN.eps,width=7cm,angle=-90}
    \hspace*{0cm}\psfig{file=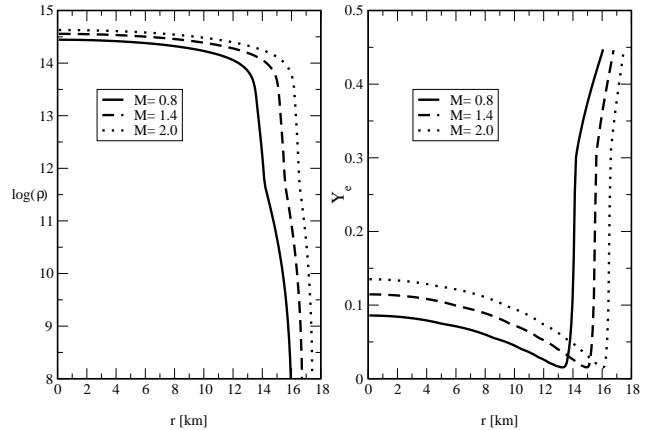,width=7cm,angle=-90}
    \caption{\label{ns_profiles} Profiles of matter density and electron 
                fraction of the initial neutron star models 
(masses in solar units).}
\end{figure}

We solve the one dimensional hydrostatic structure equations together with
the neutrino-less $\beta$-equilibrium condition to find the properties of the 
initial neutron stars. Examples of such initial profiles are shown in Fig.
\ref{ns_profiles}.
\begin{figure}
%    \hspace*{0cm}\psfig{file=Gamma_ns_1.4.ps,width=7cm,angle=-90}
    \hspace*{0cm}\psfig{file=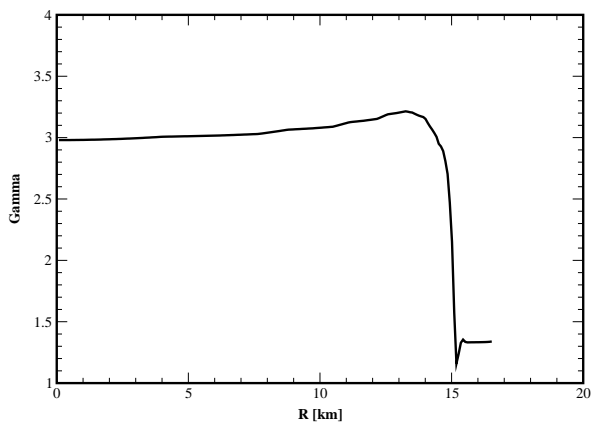,width=7cm,angle=-90}
    \caption{\label{gamma}$\tilde{\Gamma}= {\rm dln(p)/dln}(\rho)$ along 
             a neutron star profile of 1.4 M$_{\odot}$.}
\end{figure}
Note the enormous density decrease at the neutron star surface which is caused
by the stiffness of the Shen-EOS. 
This effect is more pronounced for the higher mass neutron stars, low mass 
stars possess a thicker crust, as is visible from both density and 
$Y_e$-profile.
To illustrate this stiffness we show in Fig.
\ref{gamma} the quantity $\tilde{\Gamma}= {\rm dln(p)/dln}(\rho)$ obtained by finite
differencing along a neutron star profile of 1.4 M$_{\odot}$. 
$\tilde{\Gamma}$ rises from values close to 3 in the center
to $\sim 3.2$ to drop sharply at the phase transition towards inhomogeneous 
matter and remain around 1.3 in the neutron star crust.\\
\begin{figure}
%    \hspace*{0cm}\psfig{file=Gamma_ns_1.4.ps,width=7cm,angle=-90}
    \hspace*{0cm}\psfig{file=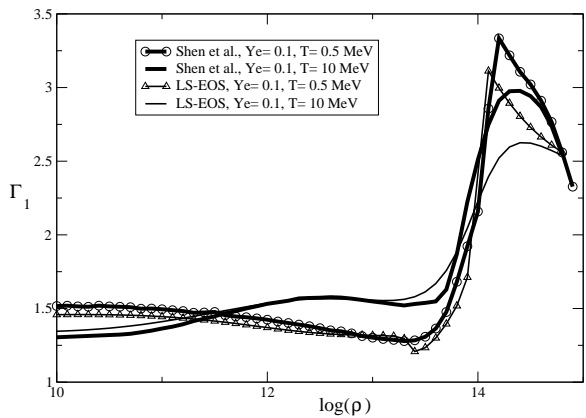,width=7cm,angle=-90}
    \caption{\label{gamma1} Shown is the variation of $\Gamma_1$ with
density (cgs-units) and temperature for a fixed electron fraction, $Y_e=0.1$,
for both our equation of state (Shen et al. plus extension) and the 
Lattimer-Swesty-EOS.}
\end{figure}
To compare once again with the LS-EOS, we plot
in Fig. \ref{gamma1} the adiabatic exponent $\Gamma_1$, given by \citep{cox68}
\begin{equation}
\Gamma_1= \frac{\rho}{p} \left(\frac{\partial p}{\partial \rho} \right)
+ T \left(\frac{\partial p}{\partial T} \right)^2  \left(\rho p \frac{\partial u}{\partial T}\right)^{-1}, \label{eq_gamma1}
\end{equation}
where $T$ is the temperature and $u$ the specific internal energy, for both
 the Shen-EOS and the LS-EOS in the relevant low-Ye regime, $Y_e= 0.1$.
In the case of the Shen-EOS the derivatives in (\ref{eq_gamma1}) have 
to be obtained by finite differencing, in the LS-EOS case they are 
available from analytical 
expressions.  For the cold case, $T=0.5 $ MeV, $\Gamma_1$ rises from values 
of $\sim1.5$ at log($\rho)= 10$ to values of $\sim 3.3$ for the Shen-EOS and
$\sim 3.1$ for the LS-EOS. Note the sharp peak at log($\rho)\approx 14$ 
which is
 present in both EOSs. In the high temperature case, $T= 10 $ MeV, the 
maximum values are slightly lower and the peaks are smeared out. 
This variation of $\Gamma_1$ with the physical conditions 
demonstrates again the difficulty of approximating a
nuclear EOS by a polytrope of fixed adiabatic exponent.\\
The electron fraction, $Y_e$, decreases from central values of around 0.1 
(higher values are encountered for more massive stars) to values as low as
$\sim 0.01$ and increases steeply again towards the surface 
(Fig. \ref{ns_profiles}). Note that
only a tiny amount of material, $\sim 1\%$ of the star's mass, is located 
in this region of increasing $Y_e$. 
\begin{table*}
\caption{Summary of the different runs. 
$a_0:$ initial separation; $\nu:$ neutrino physics; 
$T_{sim}:$ simulated duration; $M_1/M_2:$ masses in solar units; 
$\#$ part.:  total particle number}
\begin{flushleft}
\begin{tabular}{cccccccccc} \hline \noalign{\smallskip}
run & spin & ${\rm M_{1}}$ & ${\rm M_{2}}$ & \# part. & $a_{0}$ [km] 
& $\nu$ & $T_{sim}$ [ms]\\ \hline \\
A   & corot. & 1.4 & 1.4 &  207,918 & 48 & no &  10.7 \\ 
B   & corot. & 1.4 & 1.4 &1,005,582 & 48 & no &  10.8 \\
C   & irrot. & 1.4 & 1.4 &  383,470 & 48 &yes &  18.3 \\
D   & corot. & 1.4 & 1.4 &  207,918 & 48 &yes &  20.2 \\
E   & irrot. & 2.0 & 2.0 &  750,000 & 48 &yes &  12.2 \\
\end{tabular}
\end{flushleft}
\label{runs}
\end{table*}

\begin{figure*}
\caption{\label{cr3} Mass distribution in the orbital plane. 
                         The labels indicate contours of
                         log($\rho$), $\rho$ is in gcm$^{-3}$. 
		         The left column corresponds to a corotating system
                         (more than $10^6$ particles, run B) the right one to a
                         system without initial neutron star spins 
                         (more than 700 000 particles, run E). 
                         The orbital motion is counter-clockwise.}
\end{figure*}
\begin{figure*}
    \caption{\label{vel_co} Left column: velocity field and density contours 
                            in the orbital plane
                            in the corotating frame for an
                            initially corotating configuration 
                            ($> 1 \; 000 \; 000$ particles, run B). 
                            Shown is the central
                            object only (down to log($\rho$= 13.5, $\rho$ in 
                            g cm$^{-3}$). Right column: corresponding 
                            temperatures.}
\end{figure*}
\begin{figure*}
    \caption{\label{vel_ir} Left column: velocity field and density contours 
                            in the orbital plane
                            in the corotating frame for a
                            system without spins
                            ($\sim 400 \; 000$ particles, run C). 
                            Shown is the central
                            object only (down to log($\rho$= 13.5, $\rho$ in 
                            g cm$^{-3}$). Right column: corresponding 
                            temperatures.}
\end{figure*}

\subsubsection{Initial Spins and Separation}
Since the time during which the neutron stars can tidally interact is extremely
short it would need an implausibly high neutron star viscosity to lead to  
a tidal locking of the binary components during the inspiral phase 
\citep{bildsten92,kochanek92}. The spins at the moment of merger will be
negligible with respect to the orbital angular momentum and
therefore the most realistic spin configuration is close to the irrotational
case.  Thus, the most interesting configuration to be explored is the case
 without initial neutron star spins.
We also investigate corotating systems
for reasons of completeness and since it is straightforward to construct 
equilibrium configurations by damping out velocities in the corotating frame.
For details of this procedure we refer to \citet{rosswog99}.
The case of neutron stars spinning in the
direction opposite to the orbit has been explored in previous work 
\citep{ruffert96,rosswog99} and shall not be further discussed here.\\ 
The neutron stars with the corresponding spins
are then set to Keplerian orbits and are provided with radial velocities
corresponding to 
\begin{equation}
	\dot{a}_0= - \frac{64}{5} \frac{G^3}{c^5} \frac{M^2 \mu}{a^3_0},
\end{equation}
where $a_0$ is the initial separation, $M$ the total and $\mu$ the reduced mass
(e.g. \cite{shapiro83}). For further details we refer to \cite{rosswog99}.
Starting with spherical stars is obviously not an equilibrium configuration 
and will therefore result in oscillations of the stars. We have tried to reduce
these oscillations by stretching the stars according to the ellipsoidal 
approximation of \citet{lai94a} for polytropes of $\Gamma=3$ (compare Fig. 
\ref{gamma}) and correcting for the finite size effects in the corresponding 
orbital frequency. However, this did 
not improve the calculations, we therefore started the calculation with
spherical stars and Keplerian orbital frequency. At the chosen initial 
separation (see below) the tidal deformations of the stars are very small,
the height of the tidal bulge being $h\approx (R_\ast/a_0)^3 R_\ast
\approx 0.03 \; R_\ast$, where $R_\ast$ is the neutron star radius.\\
The initial separation has to be determined as a tradeoff between
computational resources and physically reasonable initial conditions.
Binary systems can become dynamically unstable (i.e. they coalesce on a
{\em dynamical} time scale) due to entirely Newtonian tidal effects
\citep{chandrasekhar75,tassoul75}.
This means that orbital decay changes abruptly from the secular orbital decay 
driven by the emission of gravitational waves to a rapid plunging of both
components towards each other on just the orbital time scale. 
The reason for this 
instability is the steeping of the effective interaction potential between
both components due to tidal effects. Tidal effects increase with the 
incompressibility of the stellar fluid (since the stars are less centrally 
condensed) and therefore the onset of this dynamical instability is a very 
sensitive function of the stiffness of the EOS, setting in at larger 
separations for a stiffer EOS. This instability has been
studied extensively in Newtonian gravity using an ellipsoidal approximation 
to the neutron star shape (\cite{lai94a} and
references therein). Relativistic effects shift the innermost stable circular
orbit, $R_{\rm ISCO}$, to larger binary separations 
\citep{baumgarte98a,baumgarte98b}.\\
We determined the binary separation where the system becomes dynamically
unstable  experimentally. 
To this end a binary configuration was relaxed in the mutual tidal field, 
set to a corotating orbit and evolved for several orbits without radial 
velocity and back-reaction force. This experiment
again underlined the enormous stiffness of the EOS which translated in a 
(for a Newtonian calculation) very large separation for the ISCO.
We found a corotating binary to be stable for an initial separation of 
$R_{dyn}= 49.5$ km, corresponding to $R_{\rm ISCO}/R_\ast \approx 3.3$. 
We therefore chose the initial 
separation for the simulation start as $a_0= 48$ km, which we regard 
to be an acceptable compromise between computational effort and physical
appropriateness.\\
The  simulations performed are summarized in Table \ref{runs}.
We use particle numbers up to $10^6$ which translates into smoothing lengths
of $\approx 0.38$ km in the initial neutron stars.

\subsection{Overall Hydrodynamic Evolution}

Since we start our simulation just inside the last stable orbit the neutron
stars approach each other very quickly and merge within roughly one orbital 
revolution. Being constantly kept back by a slowly receding centrifugal
barrier the merger itself proceeds very subsonically (keep in mind
that typical sound velocities are $\sim 0.4$ c).\\
The corotating systems (runs A, B, D) merge within approximately one orbital 
period. 
Prior to merger only a tiny lag angle develops between the axes of the 
neutron stars (first panel, left column Fig. \ref{cr3}). Note that such a lag 
angle develops even in absence of viscosity since the system is not able
to adapt fast enough to the rapidly changing tidal potential \citep{lai94b}.
Immediately after contact 
mass shedding via the outer Lagrangian points sets in which results in thick,
puffed up spiral arms (Fig. \ref{cr3}), that subsequently wrap around the 
central object to form a disk. The spiral arms show an appreciable lateral 
expansion and soon engulf the central object and the innermost high-density 
parts of the disk (last panel, left column in Fig.\ref{cr3}).\\
The irrotational configurations merge slightly quicker, the system with twice
1.4 M$_\odot$ (run C) after around three-quarters of an orbit and the system
with 2.0 M$_\odot$ (run E; Fig.\ref{cr3}) after half an orbit, due to the 
lower total angular
momentum. This faster inspiral leads to noticeably larger lag angles before
contact.
The systems start immediately to shed hot material with rather
low density from the interaction region. This is closely followed by mass
shedding via the outer Lagrangian points. The matter that is shed
by the latter
mechanism whips through the previously ejected material forming a spiral
shock which heats up the material.
The spiral pattern is much less pronounced than in the corotating case and  
becomes washed out within a few milliseconds by lateral expansion and 
a rapidly expanding disk is left behind. The material within this disk
follows eccentric trajectories and once the outward motion is reversed and 
matter starts falling back towards the hot compact remnant in the centre,
the inner parts of this torus-like structure become compressed and heat up.
Note the enormous expansion of the debris material, in all
 cases the typical diameter of the mass distribution increases
from $\sim 100$ km to $\sim 1000$ km in just a few milliseconds.\\
One may speculate that the stronger general relativistic
gravity would lead to a more compact post-merger configuration with a more
massive central object and a less massive torus around it. 
For polytropic calculations this tendency is indeed visible \citep{shibata01}.\\
In Table \ref{masses} we show the (baryonic) masses contained in the central
object and the debris material. Both regimes are discriminated on grounds of a log($\rho$)-r-plot. In several cases the transition between central object and debris is rather smooth so that it is difficult to determine the exact
mass cut. The resulting uncertainties are of the order 0.02 M$_\odot$.

\subsection{Central Object}

\begin{figure}
  \psfig{file=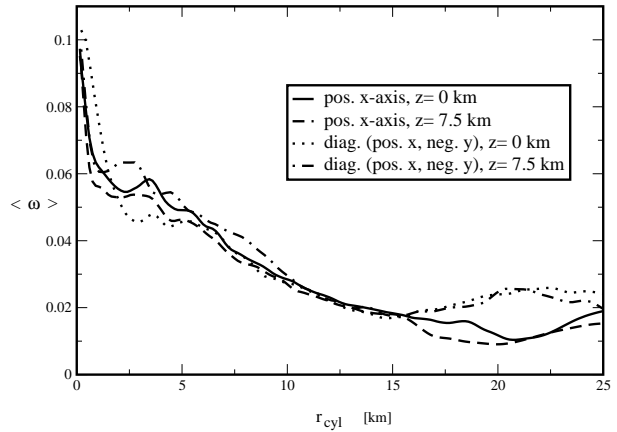,width=7cm,angle=-90}
  \caption{\label{diffrot} Differential rotation of the merged remnant I:
                           snapshot of the angular frequencies (displayed
                           in code units of $1.984 \cdot 10^5$ 1/s), 
                           of the most realistic initial configuration, run C,
                           at t= 7.686 ms (compare Fig. \ref{vel_ir}) along
	                   the positive x-axis (for two different heights)
                           and along a diagonal (positive x, negative y; again
                           two heights). The increase of $\omega$ towards the origin corresponds to the central vortex (see Fig. \ref{vel_ir}).}
\end{figure}

\begin{figure}
  \psfig{file=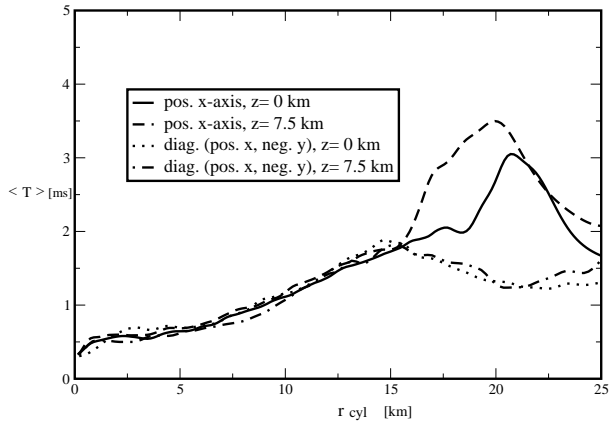,width=7cm,angle=-90}
  \caption{\label{Tdiffrot} Differential rotation of the merged remnant II:
                            periods (in ms) corresponding to Fig. \ref{diffrot}.}
\end{figure}

\begin{table}
\caption{Final rest mass distribution, ${\rm M_{co}}$ refers to the mass of
 the central object and  ${\rm M_{deb}}$ the mass of the debris material.
 $a$ denotes the stability 
parameter $Jc/GM_{co}^2$, where $J$ is the angular momentum contained in the 
central object.}
\begin{flushleft}
\begin{tabular}{cccccc} \hline \noalign{\smallskip}
run &${\rm M_{co}}$  & ${\rm M_{deb}}$ & $a$ \\ \hline \\
A   & 2.33 & 0.47 & 0.48 &  \\ 
B   & 2.35 & 0.45 & 0.49 &  \\
C   & 2.55 & 0.25 & 0.64 &\\
D   & 2.33 & 0.47 & 0.48 &\\
F   & 3.71 & 0.55 & 0.55 &\\
\end{tabular}
\end{flushleft}
\label{masses}
\end{table} 
The central objects of all performed calculations do not become axisymmetric
on the simulation time scale (Figs. \ref{cr3},\ref{vel_co},\ref{vel_ir})
and will therefore continue to emit gravitational radiation. This result is in
qualitative agreement with those from uniformly rotating stiff polytropes
($\Gamma > \Gamma_{crit} \approx 2.3$; \cite{tassoul01}). Note that contrary
to previous investigations
 \citep{davies94,rasio94} we never end up with
rigidly rotating central objects. This is an effect from the lower viscosity
due to our new artificial viscosity scheme and the high resolution of
the present calculations; the effective alpha-viscosities in the central
object are $\sim 10^{-5}$, see Section 2.1.\\
It has been pointed out previously by various authors 
\citep{ruffert96,rosswog99,rasio99,faber01} that a vortex sheet forms between 
the stars as soon as they come into contact. This is most pronounced for the 
irrotational case, since, in the frame corotating with the binary,
the stars seem to be spinning in opposite directions and therefore a 
discontinuous velocity field is encountered when the interface between
both stars is crossed.
Such a vortex sheet is known to be Kelvin-Helmholtz unstable on all wave
lengths, with the shortest modes growing fastest. As has been pointed out
previously, the shortest growing mode is determined by the 
smallest, numerically resolvable length scale. For a further discussion of 
this point we refer to the previously mentioned literature.\\
The velocity field for the initially corotating system is shown 
in column one in Fig. \ref{vel_co}.
Along the vortex sheet two vortices form which remain well-separated and which
are not dissipated until the end of the simulation.
In Fig. \ref{vel_ir} the velocity fields in a frame corotating with the 
binary system are shown (orbital plane, central object only) for the spinless
system. The vortex sheet is clearly visible in the first
panels of Figs. \ref{vel_co} and \ref{vel_ir}. 
The Kelvin-Helmholtz instability creates a set of 
vortices along the interaction interface. These merge during the further 
evolution leaving behind a rapidly, differentially rotating merger remnant.
 As will be discussed below, this  
has important consequences for the stability of the central object.\\
To further illustrate the differential rotation of the merged remnant
we show in Fig. \ref{diffrot} the SPH-smoothed values of the angular frequency
\begin{equation}
\langle \omega \rangle(\vec{r})= \sum_j \frac{v_{tan,j}}{r_{cyl,j}} \frac{m_j}{\rho_j}
W(|\vec{r}-\vec{r}_j|,h_j),
\end{equation}
where $\vec{r}$ is the point of interest (see below), $v_{tan,j}$ is the 
cylindrical tangential velocity of particle $j$,  
$r_{cyl,j}= \sqrt{x_j^2+y_j^2}$, 
$m_j$ and $\rho_j$ are mass and density and W is the used spline-based 
SPH-kernel
(e.g. \cite{monaghan92}). We only show the case of no initial spins (run C),
since it is the most relevant one (the corotating case is apart from the 
persistent vortices closer to rigid rotation). The chosen time, t= 7.686 ms,
corresponds to the last panels, column one and two in Fig. \ref{vel_ir}. 
We show 
$\langle \omega \rangle$ along the positive x-axis, both at the height of 
z= 0 and z= 7.5 km, and along a diagonal given by positive x-values and 
negative y-values,
since this line is an approximate symmetry-axis at this stage (again for 
heights of z= 0 and z= 7.5 km).
$\langle \omega \rangle$ is strongly peaked at the centre, where the central 
vortex is located and decays with increasing distance from the origin
depending on the direction and height above the 
orbital plane. The corresponding periods, 
$\langle T \rangle =2 \pi/ \langle \omega \rangle$  are shown in 
Fig. \ref{Tdiffrot}. The central 10 km of all chosen lines have periods 
below 1 ms.

\begin{figure}
  \psfig{file=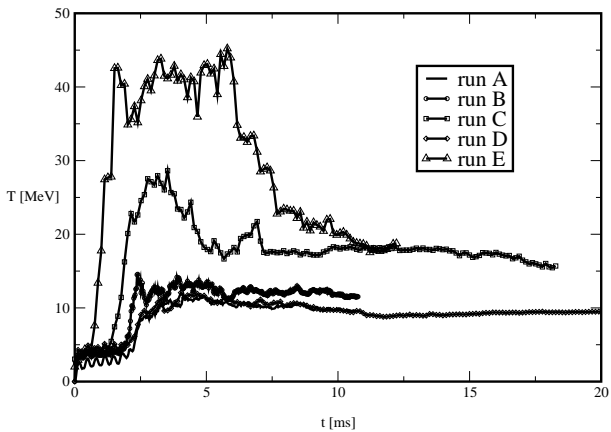,width=7.cm,angle=-90}
  \caption{\label{maxtemp} Maximum temperatures (in MeV) 
           in the merged configurations as a function of time (ms).}
\end{figure}
\begin{figure}
  \psfig{file=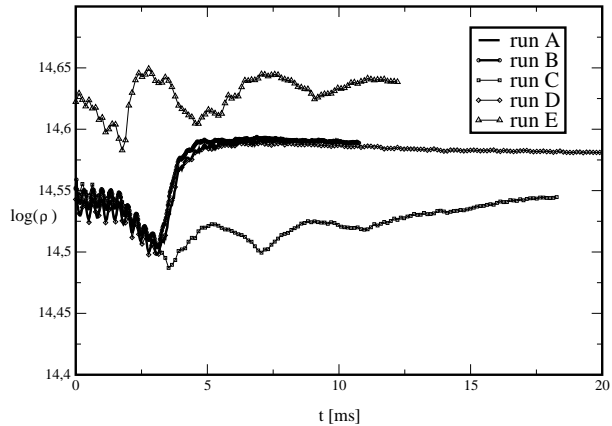,width=7.cm,angle=-90}
  \caption{\label{maxdens} Maximum densities of the different runs
                           as a function of time (measured in ms).
                           Shown is the log($\rho$) with $\rho$ in g cm$^{-3}$.}
\end{figure}
The highest temperatures of the merged 
configuration are found in the above mentioned vortices, see right
 columns Figs. \ref{vel_co} and \ref{vel_ir}.
The maximum temperatures inside the central objects are shown in
Fig. \ref{maxtemp}. 
Due to the high degeneracy of the neutron star matter oscillations due to 
imperfect initial conditions and numerical noise lead to temperatures of 
a few MeV, however only in the dense, central regions where these finite 
temperatures have no physical effect (no thermal pressure contribution at 
these densities; matter is opaque to neutrinos).
In the case of corotation the temperature hardly
rises above the level during inspiral.
Due to the larger shear motion irrotational initial conditions yield 
substantially higher maximum temperatures: up to $\sim 30$ MeV in the case
of 1.4 M$_\odot$ and $\sim 45$ MeV for our extreme case of 2.0 M$_\odot$ 
stars. These temperatures
are lower than those reported by \cite{ruffert96}. This may be an effect of 
the 
different masses used in both investigations (1.4 versus 1.6  M$_\odot$),
 the different EOS (compare also \cite{ruffert01}, Fig. 15) and different 
numerical viscosities in both codes.
\\
We find the masses of the central objects to be roughly 0.1 M$_\odot$
lower, see Table \ref{masses}, than those of previous calculations using the 
softer Lattimer-Swesty EOS \citep{rosswog99}.\\
The question whether/when the central object collapses to a black hole is
relevant for various reasons, e.g. to decide on possible mechanisms
to produce Gamma Ray Bursts (GRBs).
Most state of the art nuclear equations of state can 
support cold, non-rotating neutron stars
in $\beta$-equilibrium of $\sim 2.2$ M$_{\odot}$  gravitational mass 
\citep{akmal98}. This corresponds approximately to the baryonic masses we find here for the central objects.
Recent relativistic mean field equations of state are 
able to stabilize even heavier neutron stars (between 2.45 and 3.26 
M$_{\odot}$, \citep{chung01}).
Therefore even cold, non-rotating configurations of the masses found here
may be supported by most of these EOSs.
Additional stabilizing effects come from the finite, thermal contributions to the 
pressure. If the matter further contains non-leptonic, negative charges
such as $\Sigma^-$ hyperons, d or s quarks, then 
trapped neutrinos lead to an additional increase of the maximum mass 
\citep{prakash95}. The initial magnetic field of neutron stars, 
$B_0 \sim 10^{12}$ G, is expected to be 
 amplified in the differentially rotating remnant to enormous field strengths, 
$\sim 10^{17}$ G \citep{duncan92},
which, again, leads to a substantial increase of the maximum mass 
\citep{cardall01}.
\citet[]{stark85} have analyzed the stability of a rotating polytrope
against the collapse towards a black hole. For the investigated case of 
{\em rigid} rotation they found that rotation can stabilize the fluid
against collapse if the stability parameter $a= J c /G M^2$ is larger than 
$\approx 1$. The stability parameters $a$ of the central objects in our
calculations are $\sim 0.5$ and are shown for completeness in  Table
 \ref{masses}.
We expect the most important stabilizing effect to come from the
{\em differential} rotation of the merger remnant.
The enormous increase in the maximum mass by differential rotation has 
been demonstrated for the case of white dwarfs by  \cite{ostriker68} 
who constructed  configurations up to 4.1 M$_{\odot}$.
Recent general relativistic treatments of differentially rotating neutron 
stars \citep{baumgarte00} find that even modest degrees of differential 
rotation allow for significantly higher maximum masses than non or uniformly 
rotating configurations.\\
It may be somewhat bold to speculate from an essentially Newtonian
calculation about the stability of the remnant, but from the above line of 
argument we are tempted to conclude that the central object remains
stable at least on the simulation time scale, $\sim$ 20 ms.
Due to our ignorance of the high-density equation of state, it cannot be 
ruled out that the end product of the coalescence is a stable supermassive 
neutron star of $\sim$ 2.8 M$_{\odot}$. It seems, however, more likely
that the central object is only temporarily stabilized and once the stabilizing
effects weaken (e.g. neutrinos have diffused out after $\sim 10$ s, 
magnetic braking has damped out differential rotation) collapse to a 
black hole will occur. The time scale until collapse is difficult to 
determine, since it depends sensitively on poorly known physics and on the 
specific system parameters. We expect the most important effect to come from
the differential rotation \cite{ostriker68,baumgarte00} and therefore the
collapse time scale to be set by the time it takes the remnant to reach 
uniform rotation. Assuming the dominant 
effect to come from magnetic dipole radiation (the viscous time scale is estimated to be $\sim 10^9$ s \citep{shapiro00}), the time till  collapse
 is estimated as
\begin{eqnarray}
\tau_c \hspace*{-0.3cm} &\sim& \hspace*{-0.3cm}\frac{18 c^3 M}{5 B^2 R^4 \omega^2} \nonumber\\ 
&\sim& \hspace*{-0.3cm} 10^2 s \left(\frac{M}{2.5 M_\odot} \right) \left( \frac{10^{16} \mbox{G}}{B} 
\right)^2
\left( \frac{15 \mbox{km}}{R} \right)^4 \left( \frac{3000 \mbox{s}^{-1}}{\omega} \right)^2,
\end{eqnarray}
where $M, B$ and $R$ are mass, magnetic field and radius of the central object.

\subsection{Disk}
\begin{figure*}
    \hspace*{0cm}\psfig{file=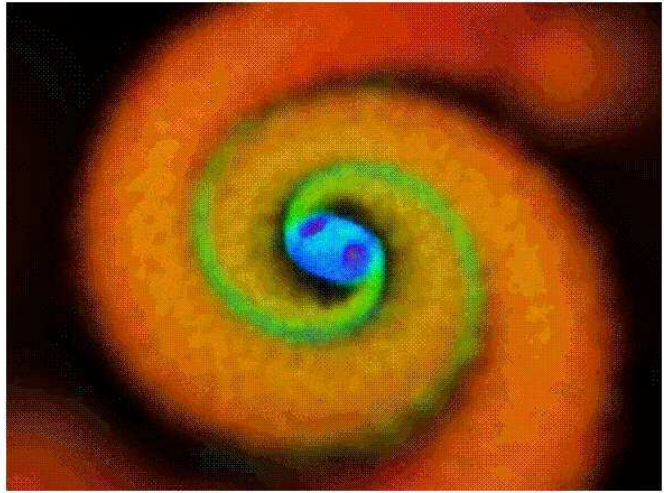,width=9cm,angle=0}\\
    \caption{\label{disk_heating} Disk heating mechanism: temperatures of the
central region shortly after the merger (t= 5.225 ms, corotation, 
$\sim 200 000$ particles, run A). The  disk is shock-heated when the
spiral arms hit supersonically due to their lateral expansion (green). The 
two hot spots (dark blue) in the central object stem from the vortices 
which formed
during the Kelvin-Helmholtz instability during the merger (see text).}
\end{figure*}
\begin{figure*}
    \caption{\label{disk} Physical properties of the disk (equatorial plane)
                          that forms  around the central object (no initial 
                          spins, run C). Shown are snapshots t= 7.434 ms 
                          (left column) and t= 11.214 ms (right column).
                          \hspace*{5cm}}
\end{figure*}
\begin{figure*}
\caption{\label{diskXZ} Vertical disk structure: density contours and velocity
fields of runs B, C and E at around 10.7 ms are shown in the left column; 
the right
column shows the temperature distribution (densities above $10^{12}$ gcm$^3$ 
are cut out to enhance the contrasts).}
\end{figure*}

\begin{figure*}
    \begin{minipage}[t]{\columnwidth}
      \hspace*{0cm}\psfig{file=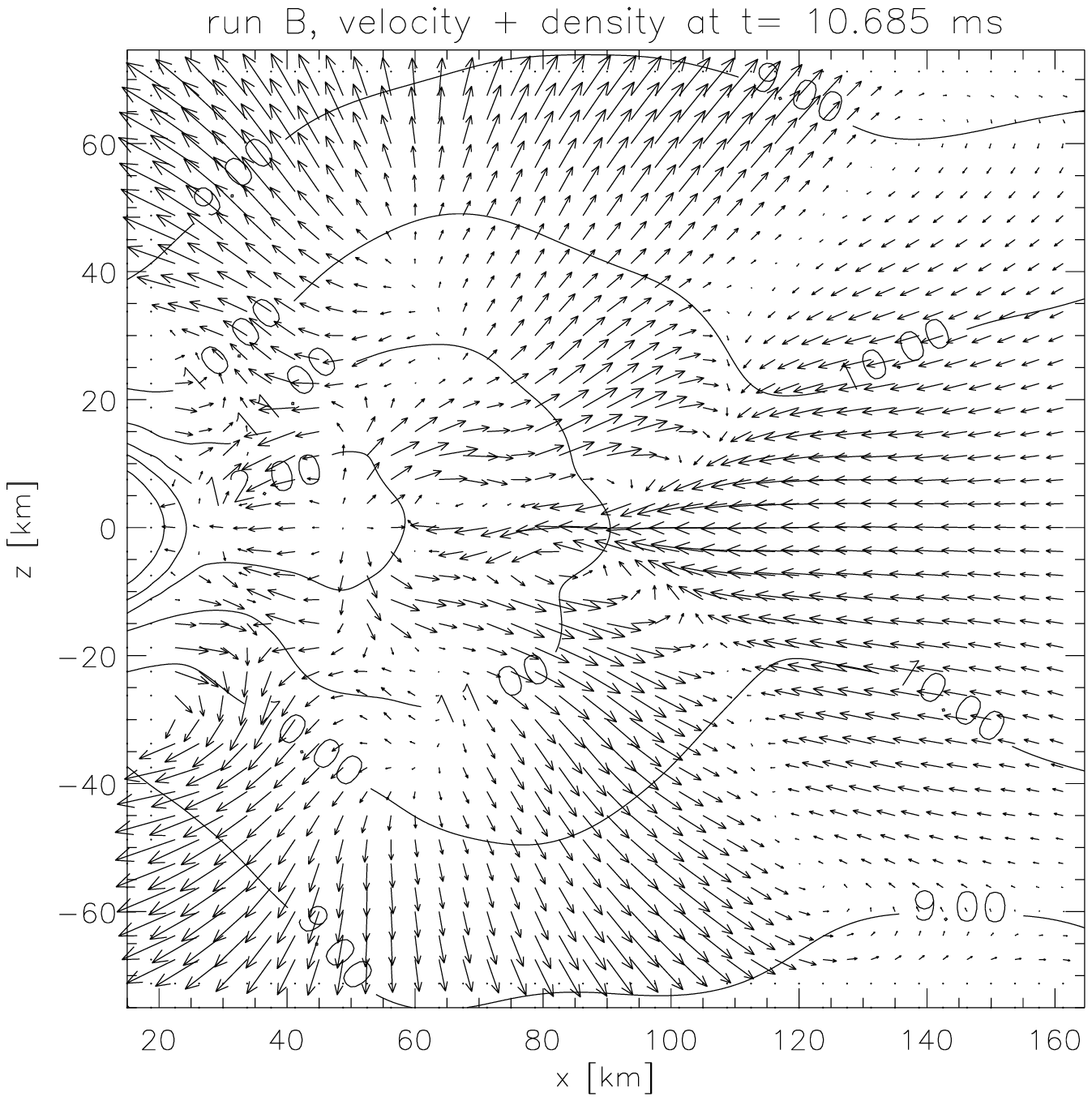,width=8.2cm,angle=0}
    \end{minipage}
    \begin{minipage}[t]{\columnwidth}
      \hspace*{0cm}\psfig{file=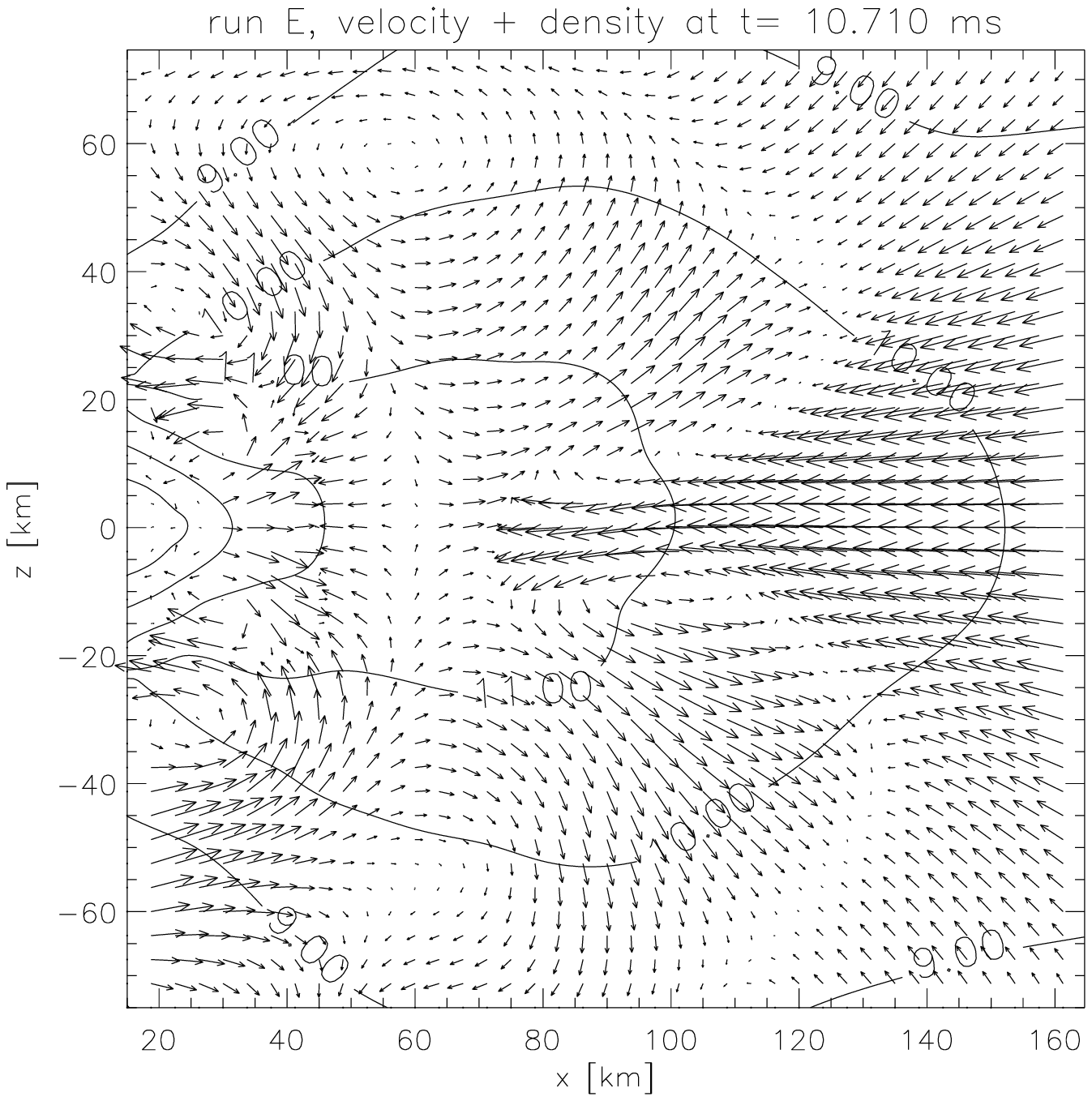,width=8.2cm,angle=0}
    \end{minipage}

\caption{\label{vel_blowup} Blow-up of the velocity fields of the inner disk 
regions of run B and E (compare to Fig. \ref{diskXZ}).}
\end{figure*}

\begin{figure}
  \psfig{file=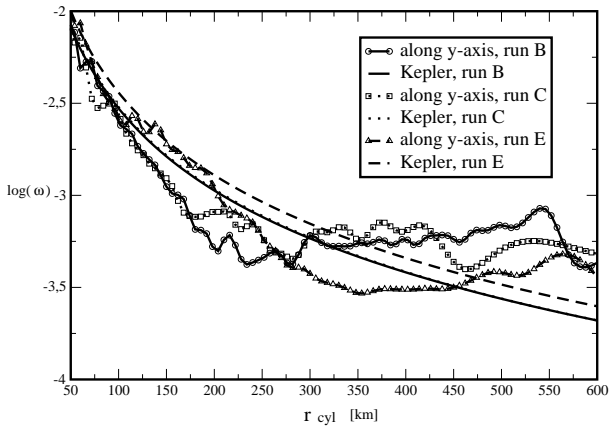,width=7.cm,angle=-90}
  \caption{\label{omega_disk} Plotted is the logarithm of the angular 
velocities (in code units of $1.984 \cdot 10^5$ 1/s) along the y-axis
as a function of the cylindrical radius for runs B (circles; t= 10.685 ms), 
C (squares; t= 10.710 ms) and E (triangles; t= 10.710 ms). 
This is compared to the corresponding Keplerian angular velocities.}
\end{figure}

Immediately after contact the two merged stars try to get rid of 
excess angular momentum by shedding mass. We refer to this debris material
as 'disk'. In all cases, a few milliseconds after contact a thick disk, with
typical heights roughly comparable to the disk radius, has formed around the
central object (see Figs. \ref{cr3}, \ref{disk} and \ref{diskXZ}).
The masses contained in the debris material are, depending on the 
initial spin, between 0.25 for run C and 0.55 M$_{\odot}$ for our extreme case,
run E.\\
Initially, as the material is being shed by the central object, the debris
material is cooled via the very fast expansion. Part of this material
is later shock-heated as subsequently ejected material runs into it. 
This is most pronounced 
in the case of initial corotation. Here the spiral arms expand laterally
due to the released nuclear energy from recombination of nucleons into nuclei
and due to the change in the adiabatic exponent and hit supersonically.
This heating mechanism is illustrated for a run A in Fig. 
\ref{disk_heating}.\\ 
Fig. \ref{disk} shows disk properties in the orbital plane for the most 
realistic, initially non-spinning, configuration. The shedded mass follows 
eccentric trajectories, resulting in a fast initial
expansion and cooling phase during the outward motion (see panel one,
left column in Fig. \ref{disk}) and a subsequent compression 
when the motion is reversed and the inner parts start to fall back towards 
the central object (see panel one, column two in Fig. \ref{disk}), 
against material that is still being shed and moving outwards.\\
To illustrate the interplay between the fluid motion, the temperature and the 
vertical disk structure, we show in Fig. \ref{diskXZ} the velocity fields 
together with density contours (left column) for run B (twice 1.4 M$_{\odot}$,
 corotation), run C (twice 1.4 M$_{\odot}$, no initial spins) and our extreme
case, run E (twice 2.0 M$_{\odot}$, no initial spins). The corresponding 
temperatures are shown in the right column. The disk shows violent convective 
motion, that can be subdivided into roughly four flow regimes:
(i) the innermost, shock-heated 'torus', that results from the collision of 
the hot matter shed from the central object and the cool (T $<$ 1 MeV) 
equatorial inflow; (ii) the cool, equatorial inflow phase; (iii) a rarefaction
regime that separates the inflow from the outflow and (vi) the cooling
 outflow from 
the disk. As the inflow hits the torus it drives a hot, vertical outflow
which leads to the thick (H$\sim$R), puffed up disk. Only the innermost, high
density material of the debris has temperatures in excess of $\approx$ 4 MeV,
the shocked torus is at around 3 MeV, the hot, high-altitude outflow with 
densities below
$10^9$ gcm$^{-3}$ has temperatures between 1 and 2 MeV, the outer parts of the 
flow close to the equatorial plane are substantially cooler (see also Fig. 
\ref{disk}, column two, panel two). In Fig. \ref{vel_blowup} we show a blow-up
of the velocity fields (run B and E) in the regime where the cool inflow 
hits the inner parts of the disk and produces the ``butterfly-shaped'' 
temperature distribution.
The temperatures found are slightly lower
than those reported by \citet{ruffert01} (up to 10 MeV).
We suspect the stiffness of the Shen-EOS in the range between $10^{12}$ and 
$10^{14}$ g cm$^{-3}$, see Fig. \ref{LS_Shen_comp}, to be partly responsible 
for this fact.\\
Despite the high temperatures we still find a substantial mass fraction
of heavy nuclei in the inner parts of the disk, which dominate the neutrino
emission. We find roughly 10 $\%$ of the disk mass in the
form of heavy nuclei (third panels in columns one and two of Fig.\ref{disk}).
The Shen-EOS yields in the inner torus regime an average nucleus with
$A \sim 80$ and $Z \sim 0.3$. We find for the ratio of the mean free path 
due to scattering off heavy nuclei, $\lambda_A$, and off free nucleons, 
$\lambda_n$, $\lambda_A / \lambda_n = n_n \sigma_n / (n_A \sigma_A) 
\approx 4 (1-X_h)/(X_h A) \approx 0.5$. Therefore {\em the scattering mean free
path in these regions is dominated by heavy nuclei}. These have not been 
accounted for in previous investigations, which assumed the
torus to be fully photo-disintegrated \citep{ruffert96}.\\
In Fig. \ref{omega_disk} we plot the logarithm of the angular velocities
as a function of the cylindrical radius for runs B, C and E and compare 
it to the corresponding Keplerian angular velocities. For runs B and C the 
innermost parts of the disk, up to $\sim 100$ km, are roughly Keplerian. 
Further out, up to $\sim 300$ km the disk rotates sub-Keplerian and the 
outermost, outflowing parts rotate faster than the corresponding Keplerian 
value. The disk of run E is roughly Keplerian out to $\sim 200$ km, drops
below the Kepler value between $\sim 200$ and $\sim 450$ km and is above 
it at larger radii.

\section{Summary and Discussion}

We have performed high-resolution 3D calculations of the last moments of the 
inspiral and the final coalescence of a neutron star binary system.
Our main motivation was to explore the impact of the, sometimes poorly known,
microphysics on the merger scenario. \\
The equations of hydrodynamics were solved using the smoothed particle 
hydrodynamics method with particle numbers of more than $10^6$
to resolve the details of the thermodynamic and nuclear evolution.
The initial neutron stars are resolved with smoothing lengths of $\approx 0.38$
km. 
We have measured the effective alpha-viscosity of our code and find for the well-resolved calculations values of $<\alpha>_{SS} \approx 6 \cdot 10^{-4}$.
The self-gravity of the fluid was treated in a Newtonian fashion, but 
forces from the emission of gravitational waves were added. \\
In these calculations  the temperature-dependent, microscopic
behaviour of the matter is modelled using a new nuclear equation of state 
whose baryonic part is described in the relativistic mean field approach 
\citep{shen98a,shen98b}. 
We have added the contributions of electron-positron
pairs (treated without any approximation) and photons to the baryonic 
component. In the low density regime the equation of state has been extended
with a gas consisting of neutrons, alpha particles, electron-positron pairs
and photons. This new EOS covers the whole parameter space in density, 
temperature and electron fraction that is relevant to the neutron star merger
scenario and therefore overcomes the restrictions inherent to previous 
calculations using a different equation of state \citep{ruffert96,rosswog99}.\\
We have also included the change in the electron fraction, $Y_e$, and 
internal energy due to the emission of neutrinos by means of a detailed
multi-flavour neutrino treatment that 
accounts carefully for the energy dependence of the neutrino reactions. 
In contrast to previous calculations scattering off heavy nuclei was included 
as a source of opacity.\\
The new equation of state is substantially stiffer than the 
previously used Lattimer-Swesty EOS, with adiabatic exponents reaching 
values well
above 3 within a  1.4 M$_{\odot}$  star. This has several important implications. First, since the stars
are less centrally condensed the binary system becomes dynamically unstable
at a larger separation. Since the inspiral behaviour changes qualitatively 
at this point from a quasi-stationary, quasi-periodic motion to a dynamic 
``plunge'' towards each other the gravitational wave signal will change 
correspondingly at this point. For a corotating system of 1.4 M$_{\odot}$ we 
find the instability to set in at a separation of $\sim$ 49 km and 
an orbital frequency of $\approx 280$ Hz corresponding to $\approx 560$ Hz
in the gravitational wave signal, well 
within the range accessible to GEO600 and LIGO \citep{schutz01}.\\
Once the neutron stars come into contact a Kelvin-Helmholtz unstable
vortex sheet forms between both stars leading to the formation of vortex rolls.
In the case of an initially corotating system two such vortices form which 
remain present and well-separated until the end of the calculation. In the
more important irrotational case, several vortices form which merge within
an orbital time scale leaving behind a differentially rotating central 
object.
It is within these vortex roles that the highest temperatures of the merged 
configuration are found, reaching peak values of 30 MeV for the case
of an irrotational system, while in the corotating case only a moderate temperature increase is observed. In a test calculation using two stars of 2.0
M$_{\odot}$ temperatures of up to 45 MeV are reached.
We find generally comparable but slightly lower temperatures than previous 
investigations that used the Lattimer-Swesty EOS.\\
The disks have a vertical height comparable to their radial extension.
We find the vertical disk structure to be determined by an equatorial inflow
of cool material that becomes shock heated in the innermost, dense disk regime.
This goes along with an outflow of hot material in the vertical direction.
In the dense inner torus around the central object we find temperatures of
3-4 MeV as compared to 5-10 MeV in the Lattimer-Sesty case. 
Despite the high temperatures we
find a substantial amount (a mass fraction of $\sim 10 \%$)
of heavy nuclei present even in the 
inner parts of the torus. This is enough to dominate the neutrino scattering
opacities which therefore play a crucial role for the neutrino transport. This 
has important consequences for the enormously temperature sensitive 
neutrino reactions and will be discussed in a future paper.\\
The masses of the central objects are roughly 0.1 M$_{\odot}$ lower
than in previous calculations where we used the Lattimer-Swesty-EOS, leading
to masses of the central object that are comparable to the maximum masses 
that most recent nuclear equations of state can support
for cold, non-rotating configurations. Since several stabilizing effects 
like thermal pressure
contributions, rotation and, most likely, magnetic fields are active during the
coalescence, we suspect the central object will remain stable on at least the 
simulation time scale of 20 ms. Probably the most important stabilisation 
comes from differential rotation. For the most probable case, where the
neutron star spins are negligible with respect to the orbital motion, the 
maximum density of the central object barely reaches the initial central
density of a single neutron star although its mass is roughly one solar mass
higher. We estimate that the central objects could be stabilized against 
collapse for as long as $\sim 10^2$ s.
It has to be stressed, however, that none of the equations of state used so
far contains particles more exotic than nucleons. The appearance of more exotic
matter in the deep cores of neutron stars is still subject to large 
uncertainties due to poorly constrained many-body interactions at highest 
densities. Therefore final conclusions on this point cannot 
be drawn.\\
If the central object indeed remains stable 
magnetic fields may wind up to reach $\sim 10^{17}$ G for a typical rotation 
period of 2 ms \citep{duncan92} and may therefore provide the conditions for 
magnetically-powered gamma ray bursts.

\vspace*{1cm}
{\bf Acknowledgements}\\
\vspace*{0.1cm}\\
It is a pleasure to thank the Leicester supercomputer team Stuart Poulton, 
Chris Rudge and Richard West for their excellent support.\\
S.R. wishes to thank M.R. Bate for useful discussions.\\
Most of the computations reported here were performed using the UK
Astrophysical Fluids Facility (UKAFF).\\
Part of this work has been performed using the University of Leicester 
Mathematical Modelling Centre's supercomputer which was purchased through 
the EPSRC strategic equipment initiative.\\
This work was supported by a PPARC Rolling Grant for Theoretical Astrophysics.
MBD gratefully acknowledges the support of the Royal Society by a URF.\\

%\input{nu_app_i}

%%%%%%%%%%%%%%%%%%%%%%%%%%%%%%%%%%%%%%%%%%%%%%%%%%%%%%%%%%%%%%%%%%%%%%%%%%%
%                                                                         %
%	                BILBIOGRAPHY                                      %
%                                                                         %
%%%%%%%%%%%%%%%%%%%%%%%%%%%%%%%%%%%%%%%%%%%%%%%%%%%%%%%%%%%%%%%%%%%%%%%%%%%
\bibliographystyle{mn2e.bst}
%\bibliography{literat}

\begin{thebibliography}{99}
\bibitem[\protect\citeauthoryear{Abramovici, {Althouse}, {Drever}, {Gursel},
  {Kawamura}, {Raab}, {Shoemaker}, {Sievers}, {Spero} \& {Thorne}}{Abramovici
  et~al.}{1992}]{abramovici92}
Abramovici A.,  {Althouse} W.~E.,  {Drever} R. W.~P.,  {Gursel} Y.,  {Kawamura}
  S.,  {Raab} F.~J.,  {Shoemaker} D.,  {Sievers} L.,  {Spero} R.~E.,
  {Thorne} K.~S.,  1992, Science, 256, 325

\bibitem[\protect\citeauthoryear{Akmal, Pandharipande \& Ravenhall}{Akmal
  et~al.}{1998}]{akmal98}
Akmal A.,  Pandharipande V.,    Ravenhall D.,  1998, Phys. Rev. C, 58, 1804

\bibitem[\protect\citeauthoryear{{Ayal}, {Piran}, {Oechslin}, {Davies} \&
  {Rosswog}}{{Ayal} et~al.}{2001}]{ayal01}
{Ayal} S.,  {Piran} T.,  {Oechslin} R.,  {Davies} M.~B.,    {Rosswog} S.,
  2001, ApJ, 550, 846

\bibitem[\protect\citeauthoryear{Balsara}{Balsara}{1995}]{balsara95}
Balsara D.,  1995, J. Comput. Phys., 121, 357

\bibitem[\protect\citeauthoryear{Baumgarte, Cook, Scheel, Shapiro \&
  Teukolsky}{Baumgarte et~al.}{1997}]{baumgarte97}
Baumgarte T.,  Cook G.,  Scheel M.,  Shapiro S.,    Teukolsky S.,  1997, Phys.
  Rev. Lett., 79, 1182

\bibitem[\protect\citeauthoryear{Baumgarte, Cook, Scheel, Shapiro \&
  Teukolsky}{Baumgarte et~al.}{1998a}]{baumgarte98a}
Baumgarte T.,  Cook G.,  Scheel M.,  Shapiro S.,    Teukolsky S.,  1998a, Phys.
  Rev. D, 57, 6181

\bibitem[\protect\citeauthoryear{Baumgarte, Cook, Scheel, Shapiro \&
  Teukolsky}{Baumgarte et~al.}{1998b}]{baumgarte98b}
Baumgarte T.,  Cook G.,  Scheel M.,  Shapiro S.,    Teukolsky S.,  1998b, Phys.
  Rev. D, 57, 7299

\bibitem[\protect\citeauthoryear{Baumgarte, Shapiro \& Shibata}{Baumgarte
  et~al.}{2000}]{baumgarte00}
Baumgarte T.,  Shapiro S.,    Shibata M.,  2000, ApJ, 528, L29

\bibitem[\protect\citeauthoryear{Benz}{Benz}{1990}]{benz90a}
Benz W.,  1990, in Buchler J.,  ed., , Numerical Modeling of Stellar
  Pulsations.
Kluwer Academic Publishers, Dordrecht, p.~269

\bibitem[\protect\citeauthoryear{Benz, Bowers, Cameron \& Press}{Benz
  et~al.}{1990}]{benz90b}
Benz W.,  Bowers R.,  Cameron A.,    Press W.,  1990, ApJ, 348, 647

\bibitem[\protect\citeauthoryear{Bildsten \& Cutler}{Bildsten \&
  Cutler}{1992}]{bildsten92}
Bildsten L.,  Cutler C.,  1992, ApJ, 400, 175

\bibitem[\protect\citeauthoryear{Bradaschia et~al.}{Bradaschia et~al.}{1990}]{bradaschia90}
Bradaschia C. et~al.,  1990, Nucl.Instrum. Methods Phys. Res. A, 289, 518

\bibitem[\protect\citeauthoryear{Cardall, Prakash \& Lattimer}{Cardall
  et~al.}{2001}]{cardall01}
Cardall C.,  Prakash M.,    Lattimer J.,  2001, ApJ, 554, 332

\bibitem[\protect\citeauthoryear{Chandrasekhar}{Chandrasekhar}{1975}]{chandras%
ekhar75}
Chandrasekhar S.,  1975, ApJ, 202, 809

\bibitem[\protect\citeauthoryear{Chung, Wang, Satiago \& Zhang}{Chung
  et~al.}{2001}]{chung01}
Chung K.,  Wang C.,  Satiago A.,    Zhang J.,  2001, astro-ph70102993, 0, 0

\bibitem[\protect\citeauthoryear{Cox \& Giuli}
{Cox \& Giuli}{1968}]{cox68}
Cox, J.P. \& Giuli, R.T. (1968), Principles of Stellar Structure, New York,
Gordon \& Breach

\bibitem[\protect\citeauthoryear{Danzmann}{Danzmann}{1997}]{danzmann95}
Danzmann K.,  1997, in of Sciences T. N. Y.~A.,  ed., Proceedings of the
  17$^{th}$ Texas Symposium on relativistic astrophysics and cosmology New York

\bibitem[\protect\citeauthoryear{Davies, Benz, Piran \& Thielemann}{Davies
  et~al.}{1994}]{davies94}
Davies M.~B.,  Benz W.,  Piran T.,    Thielemann F.-K.,  1994, ApJ, 431, 742

\bibitem[\protect\citeauthoryear{Duncan \& Thompson}{Duncan \&
  Thompson}{1992}]{duncan92}
Duncan R.,  Thompson C.,  1992, ApJ, 392, L9

\bibitem[\protect\citeauthoryear{Eichler, Livio, Piran \& Schramm}{Eichler
  et~al.}{1989}]{eichler89}
Eichler D.,  Livio M.,  Piran T.,    Schramm D.~N.,  1989, Nature, 340, 126

\bibitem[\protect\citeauthoryear{Faber \& Rasio}{Faber \&
  Rasio}{2000}]{faber00}
Faber J.,  Rasio F.,  2000, Phys.Rev. D62, p. 064012

\bibitem[\protect\citeauthoryear{Faber, Rasio \& Manor}{Faber
  et~al.}{2001}]{faber01}
Faber J.,  Rasio F.,    Manor J.,  2001, Phys.Rev. D63, p. 044012

\bibitem[\protect\citeauthoryear{Font, Dimmelmeier, Gupta \& Stergioulas}{Font
  et~al.}{2000}]{font00}
Font J.,  Dimmelmeier H.,  Gupta A.,    Stergioulas N.,  2000, astro-ph/0012477

\bibitem[\protect\citeauthoryear{Freiburghaus, Rosswog \&
  Thielemann}{Freiburghaus et~al.}{1999}]{freiburghaus99b}
Freiburghaus C.,  Rosswog S.,    Thielemann F.-K.,  1999, ApJ, 525, L121

\bibitem[\protect\citeauthoryear{Heap \& Corcoran}{Heap \&
  Corcoran}{1992}]{heap92}
Heap S.,  Corcoran M.,  1992, ApJ, 387, 340

\bibitem[\protect\citeauthoryear{J.H.~van Kerkwijk \& Zuiderwijk}{J.H.~van
  Kerkwijk \& Zuiderwijk}{1995}]{vankerkwijk95}
J.H.~van Kerkwijk J. v.~P.,  Zuiderwijk E.,  1995, A\&A, 303, 497

\bibitem[\protect\citeauthoryear{Kochanek}{Kochanek}{1992}]{kochanek92}
Kochanek C.,  1992, ApJ, 398, 234

\bibitem[\protect\citeauthoryear{Kuroda et~al.}{Kuroda et~al.}{1997}]{kuroda97}
{Kuroda} K. et~al., 1997, in Gravitational Wave
  Detection, Proceedings of the TAMA International Workshop on Gravitational
  Wave Detection held at National Women's Education Centre, Saitama, Japan, on
  12-14 November, 1996. Edited by K. Tsubono, M.-K. Fujimoto, and K. Kuroda.
  Frontiers Science Series No. 20. Universal Academy Press, Inc., 1997., p.309
  Japanese gravitational wave observatory (jgwo).
pp~309+

\bibitem[\protect\citeauthoryear{Lai}{Lai}{1994}]{lai94c}
Lai D.,  1994, MNRAS, 270, 611

\bibitem[\protect\citeauthoryear{Lai, Rasio \& Shapiro}{Lai
  et~al.}{1994a}]{lai94a}
Lai D.,  Rasio F.,    Shapiro S.,  1994a, ApJ, 420, 811

\bibitem[\protect\citeauthoryear{Lai, Rasio \& Shapiro}{Lai
  et~al.}{1994b}]{lai94b}
Lai D.,  Rasio F.,    Shapiro S.,  1994b, ApJ, 423, 344

\bibitem[\protect\citeauthoryear{Lattimer \& Schramm}{Lattimer \&
  Schramm}{1974}]{lattimer74}
Lattimer J.~M.,  Schramm D.~N.,  1974, ApJ, (Letters), 192, L145

\bibitem[\protect\citeauthoryear{Lattimer \& Schramm}{Lattimer \&
  Schramm}{1976}]{lattimer76}
Lattimer J.~M.,  Schramm D.~N.,  1976, ApJ, 210, 549

\bibitem[\protect\citeauthoryear{Lattimer \& Swesty}{Lattimer \&
  Swesty}{1991}]{lattimer91}
Lattimer J.~M.,  Swesty F.~D.,  1991, Nucl. Phys., A535, 331

\bibitem[\protect\citeauthoryear{Monaghan}{Monaghan}{1992}]{monaghan92}
Monaghan J.,  1992, Ann. Rev. Astron. Astrophys., 30, 543

\bibitem[\protect\citeauthoryear{Monaghan \& Gingold}{Monaghan \&
  Gingold}{1983}]{monaghan83}
Monaghan J.,  Gingold R.,  1983, J. Comp. Phys., 52, 374

\bibitem[\protect\citeauthoryear{Morris \& Monaghan}{Morris \&
  Monaghan}{1997}]{morris97}
Morris J.,  Monaghan J.,  1997, J. Comp. Phys., 136, 41

\bibitem[\protect\citeauthoryear{Murray}{Murray}{1995}]{murray95}
Murray, J.R., PhD thesis, Monash University (1995)

\bibitem[\protect\citeauthoryear{Oechslin, Rosswog \& Thielemann}{Oechslin
  et~al.}{2001}]{oechslin01}
Oechslin R.,  Rosswog S.,    Thielemann F.-K.,  2002, accepted for publication
in Phys. Rev. D, gr-qc/0111005 

\bibitem[\protect\citeauthoryear{Oohara \& Nakamura}{Oohara \&
  Nakamura}{1997}]{oohara97}
Oohara K.,  Nakamura T.,  1997, in Relativistic Gravitation and Gravitational
  Radiation.
Cambridge University Press, Cambridge

\bibitem[\protect\citeauthoryear{Orosz \& Kuulkers}{Orosz \&
  Kuulkers}{1999}]{orosz99}
Orosz J.,  Kuulkers E.,  1999, MNRAS, 305, 132

\bibitem[\protect\citeauthoryear{Ostriker \& Bodenheimer}{Ostriker \&
  Bodenheimer}{1968}]{ostriker68}
Ostriker J.,  Bodenheimer P.,  1968, ApJ, 151, 1089

\bibitem[\protect\citeauthoryear{Paczy$\acute{{\rm n}}$ski}{Paczy$\acute{{\rm
  n}}$ski}{1986}]{paczynski86}
Paczy$\acute{{\rm n}}$ski B.,  1986, ApJ, 308, L43

\bibitem[\protect\citeauthoryear{Prakash, Cooke \& Lattimer}{Prakash
  et~al.}{1995}]{prakash95}
Prakash M.,  Cooke J.,    Lattimer J.,  1995, Phys. Rev., D52, 661

\bibitem[\protect\citeauthoryear{Rasio \& Shapiro}{Rasio \&
  Shapiro}{1992}]{rasio92}
Rasio F.,  Shapiro S.,  1992, ApJ, 401, 226

\bibitem[\protect\citeauthoryear{Rasio \& Shapiro}{Rasio \&
  Shapiro}{1994}]{rasio94}
Rasio F.,  Shapiro S.,  1994, ApJ, 432, 242

\bibitem[\protect\citeauthoryear{Rasio \& Shapiro}{Rasio \&
  Shapiro}{1995}]{rasio95}
Rasio F.,  Shapiro S.,  1995, ApJ, 438, 887

\bibitem[\protect\citeauthoryear{Rasio \& Shapiro}{Rasio \&
  Shapiro}{1999}]{rasio99}
Rasio F.,  Shapiro S.,  1999, Class. Quant. Grav., 16, R1

\bibitem[\protect\citeauthoryear{Rosswog, Davies, Thielemann \& Piran}{Rosswog
  et~al.}{2000}]{rosswog00}
Rosswog S.,  Davies M.~B.,  Thielemann F.-K.,    Piran T.,  2000, A \&\ A, 360,
  171

\bibitem[\protect\citeauthoryear{Rosswog et al.}{Rosswog et~al.}{2002}]{rosswog01c}
Rosswog S. et al. 2002, in preparation

\bibitem[\protect\citeauthoryear{Rosswog, Liebend\"orfer, Thielemann, Davies,
  Benz \& Piran}{Rosswog et~al.}{1999}]{rosswog99}
Rosswog S.,  Liebend\"orfer M.,  Thielemann F.-K.,  Davies M.~B.,  Benz W.,
  Piran T.,  1999, A \&\ A, 341, 499

\bibitem[\protect\citeauthoryear{Ruffert \& Janka}{Ruffert \&
  Janka}{2001}]{ruffert01}
Ruffert M.,  Janka H.,  2001, A \& A, 380, 544

\bibitem[\protect\citeauthoryear{Ruffert, Janka \& Sch\"afer}{Ruffert
  et~al.}{1996}]{ruffert96}
Ruffert M.,  Janka H.,    Sch\"afer G.,  1996, A \&\ A, 311, 532

\bibitem[\protect\citeauthoryear{Ruffert, Janka, Takahashi \&
  Sch\"afer}{Ruffert et~al.}{1997}]{ruffert97a}
Ruffert M.,  Janka H.,  Takahashi K.,    Sch\"afer G.,  1997, A \&\ A, 319, 122

\bibitem[\protect\citeauthoryear{Schutz \& Ricci}{Schutz \&
  Ricci}{2001}]{schutz01}
Schutz B.,  Ricci F.,  2001, {G}ravitational {W}aves, 1. edn.
IoP, London

\bibitem[\protect\citeauthoryear{Shakura \& Sunyaev}{Shakura \& Sunyaev}{1973}]{shakura73}
Shakura, N.I. \& Sunyaev, R.A.,  1973, A \&\ A, 24, 337

\bibitem[\protect\citeauthoryear{Shapiro}{Shapiro}{2000}]{shapiro00}
Shapiro S.L., 2000, ApJ, 544, 397


\bibitem[\protect\citeauthoryear{Shapiro \& Teukolsky}{Shapiro \&
  Teukolsky}{1983}]{shapiro83}
Shapiro S.,  Teukolsky S.~A.,  1983, {B}lack {H}oles, {W}hite {D}warfs and
  {N}eutron {S}tars.
Wiley \& Sons, New York

\bibitem[\protect\citeauthoryear{Shen, Toki, Oyamatsu \& Sumiyoshi}{Shen
  et~al.}{1998a}]{shen98a}
Shen H.,  Toki H.,  Oyamatsu K.,    Sumiyoshi K.,  1998a, Nuclear Physics, A
  637, 435

\bibitem[\protect\citeauthoryear{Shen, Toki, Oyamatsu \& Sumiyoshi}{Shen
  et~al.}{1998b}]{shen98b}
Shen H.,  Toki H.,  Oyamatsu K.,    Sumiyoshi K.,  1998b, Prog. Theor. Phys.,
100, 1013

\bibitem[\protect\citeauthoryear{Shibata}{Shibata}{1999}]{shibata99}
Shibata M.,  1999, Phys. Rev. D, 60, 104052

\bibitem[\protect\citeauthoryear{Shibata \& Uryu}{Shibata \&
  Uryu}{2000}]{shibata00}
Shibata M.,  Uryu K.,  2000, Phys. Rev. D, 61, 064001

\bibitem[\protect\citeauthoryear{Shibata \& Uryu}{Shibata \&
  Uryu}{2001}]{shibata01}
Shibata M.,  Uryu K.,  2001, astro-ph/0104409

\bibitem[\protect\citeauthoryear{Stark \& Piran}{Stark \&
  Piran}{1985}]{stark85}
Stark, R.F.,  Piran T.,  1985, Phys. Rev. Lett., 55, 891

\bibitem[\protect\citeauthoryear{Sugahara \& Toki}{Sugahara \&
  Toki}{1994}]{sugahara94}
Sugahara Y.,  Toki H.,  1994, Nucl. Phys, A579, 557

\bibitem[\protect\citeauthoryear{Symbalisty \& Schramm}{Symbalisty \&
  Schramm}{1982}]{symbalisty82}
Symbalisty E. M.~D.,  Schramm D.~N.,  1982, Astrophys. Lett., 22, 143

\bibitem[\protect\citeauthoryear{Tassoul}{Tassoul}{1975}]{tassoul75}
Tassoul M.,  1975, ApJ, 202, 803

\bibitem[\protect\citeauthoryear{Tassoul}{Tassoul}{2001}]{tassoul01}
Tassoul M.,  2001, {S}tellar {R}otation.
Cambridge University Press, Cambridge

\bibitem[\protect\citeauthoryear{Taylor}{Taylor}{1994}]{taylor94}
Taylor J.,  1994, Rev. Mod. Phys., 66, 711

\bibitem[\protect\citeauthoryear{Thorsett \& Chakrabarti}{Thorsett \&
  Chakrabarti}{1999}]{thorsett99}
Thorsett S.,  Chakrabarti D.,  1999, ApJ, 512, 288

\bibitem[\protect\citeauthoryear{Timmes \& Arnett}{Timmes \&
  Arnett}{1999}]{timmes99}
Timmes F.,  Arnett D.,  1999, ApJS, 125, 277

\bibitem[\protect\citeauthoryear{Wilson \& Mathews}{Wilson \&
  Mathews}{1995}]{wilson95}
Wilson J.~R.,  Mathews G.,  1995, Phys. Rev. Lett., 75, 4161

\bibitem[\protect\citeauthoryear{Zhuge, Centrella \& McMillan}{Zhuge
  et~al.}{1994}]{zhuge94}
Zhuge X.,  Centrella J.,    McMillan S.,  1994, Phys. Rev., D50, 6247

\bibitem[\protect\citeauthoryear{Zhuge, Centrella \& McMillan}{Zhuge
  et~al.}{1996}]{zhuge96}
Zhuge X.,  Centrella J.,    McMillan S.,  1996, Phys. Rev., D54, 7261


\end{thebibliography}

\end{document}